\documentclass[acmsmall]{acmart}

\AtBeginDocument{%
  }

\usepackage{enumitem}
\usepackage{tikz}
\usepackage{xstring}

\definecolor{pone}{HTML}{FF9896}   % A1, B1, ... (light red)
\definecolor{ptwo}{HTML}{AEC7E8}   % A2, B2, ... (light blue)
\definecolor{pthree}{HTML}{98DF8A} % A3, B3, ... (light green)

\definecolor{pone}{HTML}{FF9896}   % A1, B1, ... (light red)
\definecolor{ptwo}{HTML}{AEC7E8}   % A2, B2, ... (light blue)
\definecolor{pthree}{HTML}{98DF8A} % A3, B3, ... (light green)

\DeclareRobustCommand*{\circledbox}[3]{\tikz[baseline=(char.base)]{%
  \node[shape=circle, draw=black, fill=#1, line width=0.5pt, inner sep=0.5pt,
        minimum size=1em, font=\scriptsize, text=#2] (char) {#3};}}

\DeclareRobustCommand*{\circled}[1]{%
  \IfStrEqCase{#1}{%
    {I}{\circledbox{black}{white}{#1}}{II}{\circledbox{black}{white}{#1}}%
    {III}{\circledbox{black}{white}{#1}}{IV}{\circledbox{black}{white}{#1}}%
    {V}{\circledbox{black}{white}{#1}}{VI}{\circledbox{black}{white}{#1}}%
    {VII}{\circledbox{black}{white}{#1}}{VIII}{\circledbox{black}{white}{#1}}%
    {IX}{\circledbox{black}{white}{#1}}{X}{\circledbox{black}{white}{#1}}}%
    [\circlednonroman{#1}]}
\DeclareRobustCommand*{\circlednonroman}[1]{%
  \StrRight{#1}{1}[\circledlast]%
  \IfStrEqCase{\circledlast}{%
    {1}{\circledbox{pone}{black}{#1}}%
    {2}{\circledbox{ptwo}{black}{#1}}%
    {3}{\circledbox{pthree}{black}{#1}}}%
    [\circledbox{white}{black}{#1}]}
\DeclareRobustCommand*{\plaincircled}[1]{\circledbox{white}{black}{#1}}
\usepackage[normalem]{ulem}

\begin{document}

\setcopyright{cc}
\setcctype{by}
\acmJournal{PACMHCI}
\acmYear{2027} \acmVolume{11} \acmNumber{1} \acmArticle{GROUP19}
\acmMonth{1} \acmDOI{10.1145/3859405}

\title[From Artifacts to Coordination]{From Artifacts to Coordination: Exploring How Shared Canvases Drive Collaboration during Group Ideation}

\author{Saramsh Gautam}
\email{sgauta4@lsu.edu}
\affiliation{%
  \institution{Louisiana State University}
  \department{Division of Computer Science and Engineering, College of Engineering}
  \city{Baton Rouge}
  \state{Louisiana}
  \country{USA}
}

\author{Andrew M. Webb}
\email{andrewwebb@lsu.edu}
\affiliation{%
  \institution{Louisiana State University}
  \department{Division of Computer Science and Engineering, College of Engineering}
  \city{Baton Rouge}
  \state{Louisiana}
  \country{USA}
}

\author{Mahmood Jasim}
\email{mjasim@lsu.edu}
\affiliation{%
  \institution{Louisiana State University}
  \department{Division of Computer Science and Engineering, College of Engineering}
  \city{Baton Rouge}
  \state{Louisiana}
  \country{USA}
}

\renewcommand{\shortauthors}{Saramsh Gautam, Andrew M. Webb, and Mahmood Jasim}

\begin{abstract}
% Recent years have seen a surge in commercial and research tools that support collaborative ideation through shared canvases. 
% While these canvases offer flexible workspaces for collaboration, how collaborators leverage such canvases to iteratively develop ideas by externalizing thoughts, organizing artifacts, and coordinating tasks remains underexplored.
% To address this gap, we conducted an exploratory study with seven triads collaboratively planning a travel itinerary using an infinite-canvas technology probe. 
% Drawing on participant interviews and video analysis of their canvas interactions, we observed that collaborators contributed in parallel, placing and arranging artifacts freely without a predefined structure. 
% We found that this spatial organization allowed collaborators to coordinate implicitly, made less vocal participants' contributions visible, helped distinguish private exploratory regions from shared working areas, and supported the emergence of distinct participant roles through repeated patterns of contribution.
% We discuss design implications for canvas-based ideation tools, including surfacing unnoticed contributions, supporting emergent roles without fixing assignments, and keeping spatial cues legible for AI integration in collaborative canvases.
Recent years have seen a surge in tools that support collaborative ideation through shared canvases. While these canvases offer flexible workspaces, how collaborators leverage them to iteratively develop ideas by externalizing thoughts, organizing artifacts, and coordinating tasks remains underexplored. To address this gap, we conducted an exploratory study with seven triads planning a travel itinerary using an infinite-canvas technology probe. Drawing on participant interviews and video analysis of their canvas interactions, we observed that collaborators contributed in parallel and arranged artifacts freely without a predefined structure. This spatial organization allowed collaborators to coordinate implicitly, made less vocal participants' contributions visible, helped distinguish private exploratory regions from shared working areas, and supported the emergence of distinct roles through repeated patterns of contribution. We discuss design implications for canvas-based ideation tools, including surfacing unnoticed contributions, supporting emergent roles without formalizing assignments, and keeping spatial cues legible for AI integration in collaborative canvases.
\end{abstract}



% \keywords{Do, Not, Use, This, Code, Put, the, Correct, Terms, for, Your, Paper}


\received{31 March 2026}
\received[revised]{17 July 2026}
\received[accepted]{17 September 2026}

\maketitle

\section{Introduction}

Collaborative ideation involves groups of people generating, sharing, and discussing ideas together.
The effectiveness of the ideation process often depends on how ideas are represented~\cite{fischer2005social, goldschmidt1994visual, sawyer2017group}. 
Representations that make ideas visible allow groups to build on each other's work, while persistent representations allow them to revisit contributions and compare ideas over time~\cite{sawyer2017group, transcending2000ernesto}.
To support these processes, groups rely on various digital tools, including video conferencing tools such as Zoom~\cite{zoom2024} and Microsoft Teams~\cite{teams}, messaging tools such as Slack~\cite{slack} and Discord~\cite{discord}, and word processing tools such as Google Docs~\cite{googledocs}.
However, these tools are disconnected from one another, which scatters ideas across separate spaces and makes it difficult for groups to maintain a shared, persistent view of contributions~\cite{dourish1992awareness}. This creates challenges in maintaining awareness and coordination---key components of collaborative ideation~\cite{dourish1992awareness,gutwin2002descriptive}.
For instance, Google Docs provides limited visibility of user intention until after actions are taken, making it difficult for groups to maintain shared awareness and coordinate activities~\cite{jung2017googledocs}.

In recent years, infinite canvases, such as Miro~\footnote{\href{https://miro.com/}{https://miro.com/}}, have received attention for providing an integrated, free-form space where collaborators can express ideas, arrange and revise contributions, and work together on multiple ideas within the same shared space~\cite{frich2019mapping, shneiderman2006creativity, macneil2023freeformtemplates, hamilton:2018}.
Such canvases have the potential to address the disconnection of ideas by bringing contributions into one persistent space that all collaborators can access~\cite{gronbaek2024blended}. 
They also support the curation of heterogeneous content, letting collaborators collect, place, and annotate different media types --- text, images, and video --- within the same space~\cite{webb2016free, kerne:2017}.
Prior work has explored how participants collaborate in co-located tabletop surfaces~\cite{Scott2004Territoriality}, large shared displays~\cite{Kirshenbaum2021Traces}, and domain-specific professional tools~\cite{Frich2019Strategies}.
However, collaboration in remote canvases differs from co-located tools in one key aspect relevant to ideation: participants are not co-present, so they cannot rely on gesture, gaze, and proximity to understand intent and coordinate actions~\cite{kraut2002use, Dangelo2016Gazed}.
% and the canvas can hold heterogeneous and incomplete idea representations.
Hence, how remote collaborators use the open layout and persistent artifacts of infinite canvases to coordinate ideation activities, maintain awareness of others' work, and develop shared work over time remains underexplored.
In this work, we explore the following research questions:
\begin{itemize}[leftmargin=2em]
    \item RQ1: How do remote collaborators express and build on each other's ideas in infinite-canvas workspaces?
    \item RQ2: How does spatial flexibility influence coordination and interpretation during collaborative ideation?
\end{itemize}

To investigate how remote collaborators engage in collaborative ideation using an infinite canvas, we conducted an exploratory study with seven triads.
Each triad engaged in a remote travel-planning session using an infinite-canvas technology probe.
The canvas allowed participants to collect, organize, and annotate multimedia artifacts, including text, images, sketches, audio, video, and shapes.
We asked each triad to collaboratively plan a travel itinerary for a randomly assigned continent, requiring participants to gather and organize information into a shared plan by the end of the session.
We investigated how collaborators expressed ideas, arranged artifacts, revisited prior contributions, and developed shared plans through sustained interaction with the canvas.
We also conducted semi-structured group interviews to capture participants’ reflections on their collaboration process and their experience working on a shared canvas.

We conducted video-based interaction analysis~\cite{jordan1995interaction} of each collaborative planning session, complemented by thematic analysis of post-task semi-structured interviews.
We found that spatial actions, such as placement, movement, and grouping of artifacts, served as an implicit communication resource that allowed participants to signal intent and priority without explicit verbal exchange, and that gave less vocal collaborators a visible way to contribute.
Over the course of collaboration, this spatial organization gave rise to distinct regions on the canvas, with central areas reserved for shared, finalized ideas and peripheral areas supporting more private, exploratory work.
Our analysis also shows how roles emerged through participants' organizational work on a remote infinite canvas, with responsibilities derived from their work instead of explicit negotiation, extending what is known about implicit coordination and emergent roles in other collaborative settings~\cite{gutwin1996workspace,welser2011finding}.
Based on our findings, we discuss design implications for canvas-based collaborative ideation tools that better support coordination, awareness, and role emergence in remote settings.
We contribute:
 \begin{itemize}[leftmargin=2em]
    \item An empirical account of how small groups of remote collaborators express ideas, organize artifacts, and develop shared plans through sustained interaction on a multimodal infinite canvas;
    \item Insights into how artifact organization served as a coordination and participation mechanism through which remote collaborators signaled intent, contributed without speaking, and made emerging divisions of labor visible;
    \item Design implications for supporting collaborative coordination and awareness within canvas-based platforms.
\end{itemize}

\section{Related Work}

In this section, we situate our work within prior research on collaborative ideation, digital collaboration workspaces, and canvas environments.
We review how ideation unfolds through iterative interaction with externalized artifacts, how existing collaboration tools support that process, and how spatial organization has been studied in co-located and domain-specific settings.

\subsection{Collaborative Ideation as Iterative and Externalized Work}
During collaborative ideation, participants iteratively develop, reinterpret, and extend each other's contributions through interaction over time~\cite{fischer2005social, sawyer2017group}. 
Prior work has shown that ideas in collaborative settings often emerge gradually, with early contributions functioning as prompts, fragments, or starting points that are revisited and transformed as collaboration unfolds~\cite{fischer2005social, sawyer2017group, transcending2000ernesto, warr2007externalizations}. 
From this perspective, collaborative ideation goes beyond generating novel suggestions towards navigating and sustaining ideation through revisiting and building on evolving contributions.

Research on creative practice, visual thinking, and external cognition has emphasized the importance of external representations in supporting the creative ideation process~\cite{hollan2000distributed, kirsh2010thinking, buxton2010sketching, sterman2023kaleidoscope}. 
Sketches, notes, diagrams, and other intermediate artifacts allow people to express incomplete thoughts, compare alternatives, and offload emerging ideas into forms that can be revisited and revised~\cite{arnheim2023visual, goldschmidt1994visual, kirsh2010thinking, buxton2010sketching}. 
These representations enable idea development by making tentative thoughts available for reflection, reinterpretation, and further elaboration~\cite{hollan2000distributed, perry2003distributed, macneil2023freeformtemplates}. 
In collaborative settings, such externalizations also become shared referents that participants can point to, modify, and discuss together, helping groups construct shared understanding through ongoing interaction with evolving materials~\cite{transcending2000ernesto, feng2023UX, wang2024meetingbridges}.

Collaborative ideation often involves incomplete ideas and ongoing changes which can benefit from environments that allow collaborators to put ideas down, revise them, and reorganize them without having to commit to a fixed structure~\cite{shneiderman2006creativity, buxton2010sketching, hamilton:2018}. 
Therefore, the design of ideation environments matters for how ideas are recorded and how participants engage with each other through shared idea representations. 
This underscores the importance of examining how collaborative systems support the externalization, visibility, and continued development of ideas in practice.

\subsection{Workspaces in Collaboration and Ideation Support Tools}

Collaborative work is often distributed across systems that support communication, coordination, and shared production across time and space~\cite{olson2000distance, grudin1994groupware, gutwin1996workspace}. 
Messaging platforms such as Slack~\cite{slack}, Discord~\cite{discord}, and Microsoft Teams~\cite{teams} help organize discussion chronologically supporting idea exchange and coordination, while shared documents such as Google Docs~\cite{googledocs} and Microsoft Word~\cite{microsoftword} support co-authoring within a linear textual form. 
Such tools privilege sequential, text-centric exchange, in contrast to the flexible externalization, spatial comparison, and simultaneous engagement that ideation requires~\cite{shneiderman2006creativity, frich2019mapping}.
Dedicated collaborative workspaces address this gap by giving teams a shared, persistent space in which to place, arrange, and revise material as a group~\cite{gutwin1996workspace, hamilton:2018}.

Sustaining parallel and evolving work on such a workspace requires mechanisms that let collaborators track each other's activity and the changing state of shared material~\cite{gutwin1996workspace, gutwin2002descriptive, tan2024rcp, saffo2023aware}.
Prior work in computer-supported cooperative work and collaborative ideation has explored how such workspaces support coordination, looking at practices including the organization of shared material on tabletops~\cite{Scott2004Territoriality}, and within domain specific software~\cite{Frich2019Strategies}. Also some works examine how contributions accumulate and persist on physical whiteboards~\cite{cherbuni2007whiteboard} and large shared displays~\cite{Kirshenbaum2021Traces}.
Recent work extends this to digital whiteboards  examining collaboration dynamics in hybrid teams around a shared canvas~\cite{neumayr2022territoriality, neumayr2025hybrid}.

However, this body of work studies coordination where collaborators share physical proximity or hybrid setup~\cite{neumayr2022territoriality, gronbaek2024blended}.
Fully remote collaborations lack physical proximity and cues such as gesture and gaze that supports coordination in co-located settings~\cite{kraut2002use, Dangelo2016Gazed}.
This motivates closer attention to how workspace itself shapes expression, coordination, and the interpretation of evolving collaborative work.

% \subsection{Open Canvas Workspaces, Spatial Organization, and Artifact-Mediated Coordination}
\subsection{Canvas-based Workspaces and Spatial Organization}

Canvas workspaces, such as Miro~\footnote{\href{https://miro.com/}{https://miro.com/}} and Figjam~\cite{figjam}, offer a more flexible environment by removing the boundaries of fixed surface that constrains collaborators to work within the edges of tabletops, displays, or pages~\cite{macneil2023freeformtemplates, hamilton:2018}.
These environments allow collaborators to pan and zoom freely, and place contribution wherever they choose across an unbounded space.
This flexibility suits collaborative ideation process, since it allows collaborators to externalize partial thoughts, explore alternatives, and revisit evolving material without committing to a structure early on~\cite{frich2019mapping, shneiderman2006creativity, macneil2023freeformtemplates}.

Research on distributed cognition shows that thinking and coordination are distributed across people, tools, and representations~\cite{hollan2000distributed, aragon2011collaborative}. 
From this perspective, collaborative ideation unfolds as members of a group externalize and act on shared artifacts~\cite{kirsh2010thinking}.
On a canvas, this externalization is inherently spatial because every artifact a collaborator places also carries a position relative to previously added contributions. 
While prior work has established the importance of shared artifacts, spatial organization, and workspace awareness, less is known about how these dynamics unfold in remote canvas-based environments during collaborative ideation, where multiple contributions can emerge, shift, and be reorganized across a shared space over time.

\section{Methodology}
We studied seven triads using an infinite-canvas technology probe to examine how participants expressed and discussed their ideas, organized curated artifacts, and coordinated ideation activities while collaboratively planning a travel itinerary. 
Each collaborative session concluded with a group interview to collect participants' reflections on their experience.

\subsection{Infinite Canvas Technology Probe}
\begin{figure*}[tb]
    \centering
    \includegraphics[width=\textwidth]{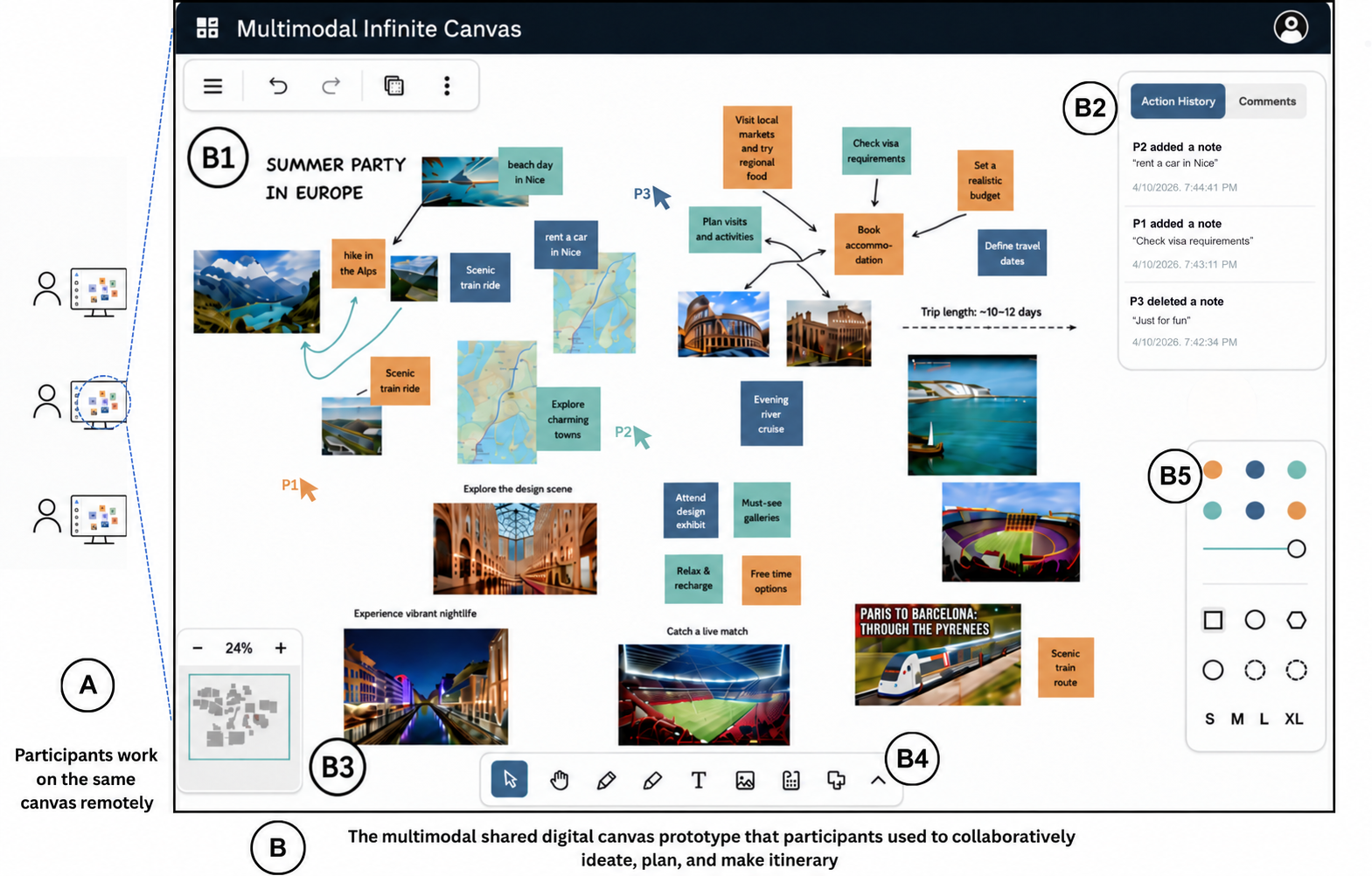}
    \caption{Overview of the technology probe. 
    % and procedure. 
    \circled{A} Three participants join remotely via Zoom and work on the shared canvas. 
    \circled{B} The multimodal, infinite-canvas technology probe. \plaincircled{B1} shared canvas space with multimodal input support, \plaincircled{B2} usage log panel, \plaincircled{B3} minimap for navigation, \circled{B4} toolbar with tools for multimodal input, \circled{B5} color and style palette to edit the contents.
    % C: Participants independently add vacation-related ideas to the canvas. 
    % D: Participants collaboratively discuss, refine, and organize ideas. 
    % E: Groups present and explain their finalized plans. 
    % F: Semi-structured interview capturing reflections on the task and their collaboration process.
    }
    \Description{A figure showing the study setup and procedure. Panel A shows three participants joining a remote session through Zoom. Panel B shows the collaborative technology probe interface, including a shared infinite canvas for multimodal inputs, a usage log panel, and a minimap for navigation. 
    % Panel C shows participants independently adding vacation-related ideas to the canvas. Panel D shows the group collaboratively discussing, refining, and organizing ideas on the shared canvas. Panel E shows the group presenting and explaining their finalized plan. Panel F shows a semi-structured interview used to capture reflections on the task and collaboration process.
    }
    \label{fig:procedure}
\end{figure*}

We built a web-based technology probe~\cite{hutchinson2003techprobe} to support multimodal, infinite-canvas collaboration. We chose to build a custom probe rather than using a commercial tool such as Miro to enable detailed interaction logging at a level that commercial platforms do not support, and to provide a simplified environment that would not overwhelm participants with features irrelevant to our research questions.
The probe was built using TLDraw~\footnote{\href{https://tldraw.dev/}{https://tldraw.dev/}}, an open-source infinite-canvas library.
It supported free-form web curation, allowing multiple users to simultaneously collect and spatially organize artifacts across various media types, including text, images, sketches, audio, video, and shapes (Fig.~\ref{fig:procedure}, B1).
Participants could freely place artifacts anywhere on the canvas, move and reposition them at any point, resize, rotate, and layer contents, and group related artifacts together.
Participants could also annotate existing content, draw freehand, and rearrange artifacts without any fixed layout constraints.
A usage log panel allowed participants to review a chronological record of interactions by all group members throughout the session (Fig.~\ref{fig:procedure}, B2).
To navigate the canvas, participants could freely pan and zoom their individual views.
We included a minimap with a persistent, holistic overview of the canvas 
(Fig.~\ref{fig:procedure}, B3). 

\subsection{Participants}
We advertised our study through academic mailing lists, flyers, and posts on LinkedIn and Bluesky.
We selected participants at random from a pool of interested individuals who met our inclusion criteria---an adult aged 18 to 65 and fluent in English. 
We had a total of 21 participants, with reported age ranges of 18--24 (5/21), 25--34 (15/21), and 35--44 (1/21). 
Ten participants used he/him pronouns, ten used she/her, and one used they/them pronouns.
Participants came from a range of educational backgrounds, including high school diploma or equivalent (2), bachelor's degree (9), master's degree (5), and doctorate (PhD) degree (5). 
Most participants worked in academic settings (17), while others came from corporate or industry settings (3) or were self-employed or freelance (1). 
All participants had prior experience with collaborative tools in some form. 8/21 participants had experiences with collaborating using physical artifacts, including whiteboards and sticky notes. Others used digital collaborative tools (13/21 participants), such as Google Docs~\cite{googledocs}, Notion~\cite{notion2024}, Zoom~\cite{zoom2024}, Slack~\cite{slack}, Jamboard~\footnote{\href{https://edu.google.com/products/jamboard/}{https://edu.google.com/products/jamboard/}}, or Miro~\footnote{\href{https://miro.com/}{https://miro.com/}}.
Participants' backgrounds included computer science, engineering, mathematics, and business. 
From the 21 participants, we formed seven triads (A--G) at random, balancing for their availability and background diversity, which we achieved by ensuring that the groups had no more than one participant from the same background. 
Additionally, we verified that no members of the triad knew each other prior to the study.

\begin{figure*}[tb]
    \centering
    \includegraphics[width=\textwidth]{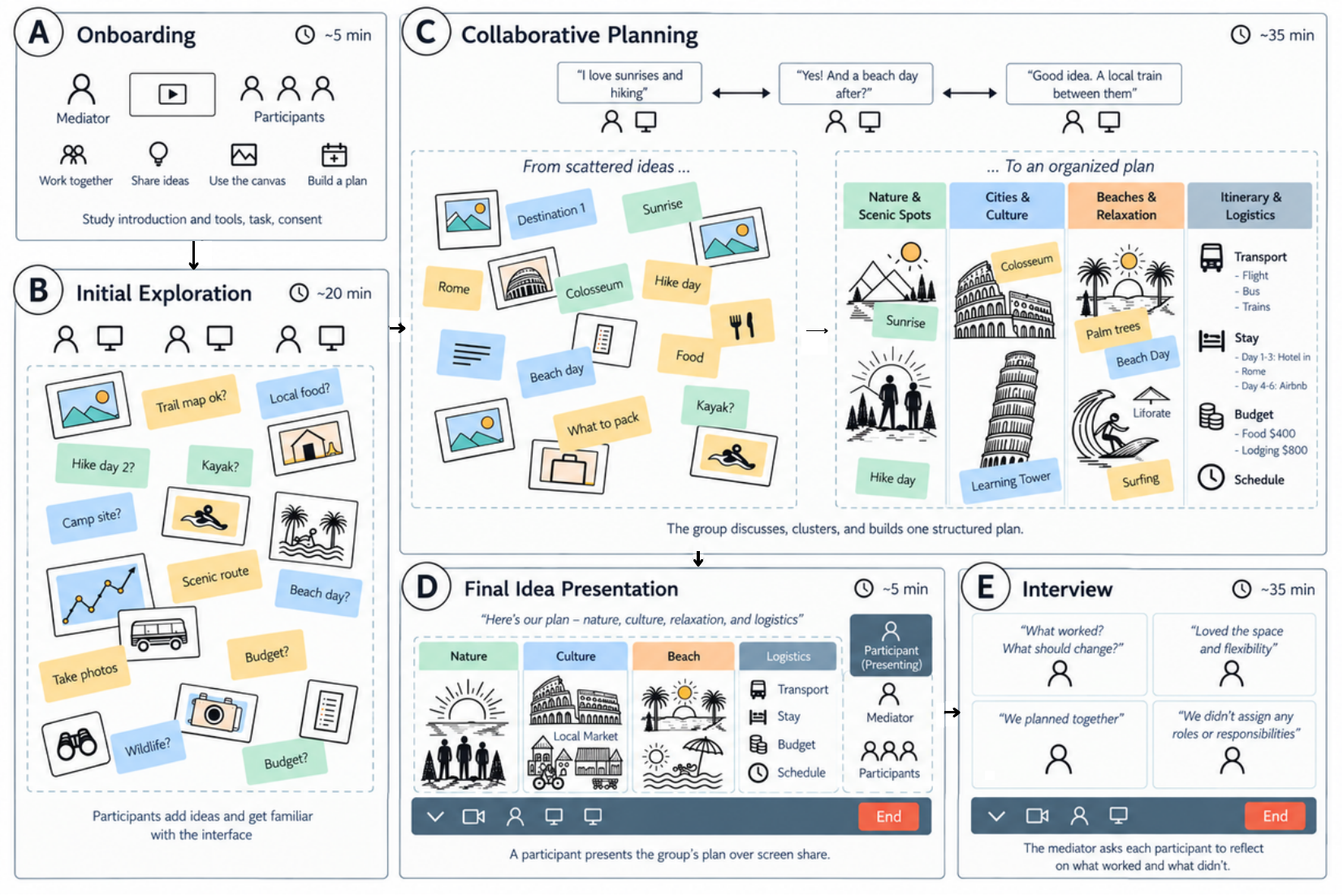}
    \caption{Overview of the study phases. 
    \circled{A} The researcher introduced the task and the shared workspace. 
    \circled{B} Participants engage in initial exploration by becoming familiar with the shared canvas and contributing early travel-related ideas. 
    \circled{C} Participants collaboratively add ideas as images, text, notes, and videos. They discuss, refine, compare, and spatially organize ideas and made an itinerary of the travel plans.  
    \circled{D} A member of each group presented their final itinerary using the canvas as a shared reference. 
    \circled{E} Participants take part in a semi-structured interview reflecting on their collaboration process and use of the workspace.}
    \Description{A five-panel figure showing the study procedure. Panel A shows the onboarding phase, where researchers introduce the task and shared digital workspace. Panel B shows the initial exploration phase, where participants explore the shared canvas and add early travel-related ideas. Panel C shows the collaborative planning phase, where participants discuss, refine, compare, and spatially organize ideas into a shared itinerary. Panel D shows the final presentation phase, where groups present their finalized travel itinerary using the canvas. Panel E shows the semi-structured interview phase, where participants reflect on collaboration and workspace use. A timeline underneath the panels indicates a total session time of approximately 95 to 100 minutes.}
    \label{fig:study-phases}
\end{figure*}

\subsection{Procedure and Task}

We conducted all study sessions remotely via Zoom, and each session followed a sequence of phases, illustrated in Fig. ~\ref{fig:study-phases}. 
The session began with a brief onboarding (M = 5 min, range = 2--9 min; Fig.~\ref{fig:study-phases}A), during which the researcher introduced the canvas probe through a demo video and answered clarifying questions.
We assigned a randomly selected continent to each group and instructed them to use the shared canvas to collaboratively plan a vacation itinerary, gathering and organizing ideas around destinations, activities, and logistics.
Participants were first asked to work individually on the shared canvas to become familiar with it and add initial ideas (M = 20 min, range = 9–31 min; Fig.~\ref{fig:study-phases}B). 
The facilitator then prompted each group to begin collaborative planning.
This individual-first structure follows established best practice in collaborative ideation research, which suggests that individual generation before group work reduces fixation effects, minimizes production blocking, and supports more diverse contributions~\cite{diehl1987productivity, kohn:2011, nijstad2003turntaking}.
The collaborative planning phase concluded when the group considered their plan complete (M = 35 min, range = 26--44 min; Fig.~\ref{fig:study-phases}C).
During this phase, participants discussed, refined, and spatially organized their ideas into a shared plan.
Each group then presented their finalized itinerary using the canvas as a reference (M = 5 min, range = 2--8 min; Fig.~\ref{fig:study-phases}D).
The session concluded with a semi-structured group interview (M = 35 min, range = 23--47 min; Fig.~\ref{fig:study-phases}E), in which participants reflected on their collaboration process and experience with the probe, with follow-up questions tailored to specific behaviors observed during each session.

\begin{table*}[t!]
\small
\centering
\caption{Summary of the three themes derived from the study. The table lists each theme, with representative codes and a brief description, showing how participants used shared artifacts to implicitly coordinate, develop spatial structure over time, and differentiate collaborative roles through repeated patterns of contribution.}
\Description{Table summarizing three themes derived from the exploratory study. It lists each theme, along with representative codes and a description.}
\label{tab:themes}
\begin{tabular}{p{3cm} p{5cm} p{5cm}}
\toprule
\textbf{Theme} & \textbf{Codes} & \textbf{Description} \\
\midrule

Implicit Coordination Through Artifact Placement
&
Captioning images; visual anchoring and labeling; spatial rearrangement; overlapping entry, proximity-based entry; layout reused by others
&
Collaborators used the placement, grouping, and elaboration of artifacts to communicate intent and indicate relationships, allowing coordination to emerge through the canvas itself. \\

\hline

Spatial Structure Emerges Through Use
&
Scattered early layout; parallel idea threads; clustering; central shared regions; peripheral exploratory regions; personal and common spaces
&
Even without predefined structure, the canvas became organized over time into central and peripheral regions taking on different functions as ideas were revisited, clustered, and negotiated. \\

\hline

Roles Emerge Through Artifact Work
&
Organizing canvas elements; layout maintenance; curating content; primary canvas editor; idea presentation; plan awareness; task assignment
&
Roles were not usually assigned explicitly, but emerged through repeated patterns of contribution, with organizing work making task efficacy and divisions of labor visible. \\

\bottomrule
\end{tabular}
\end{table*}

\subsection{Data Collection and Analysis}
We collected data from three sources: a pre-questionnaire that captured demographic information; 600 minutes of screen recordings along with mediators' observation notes of group behavior and interaction flow; and post-task interviews that collected participants' feedback and reflections.

We employed a multi-method analysis strategy, combining video-based interaction analysis~\cite{jordan1995interaction} and thematic analysis~\cite{braun2006using}.
For the video analysis, two coders independently reviewed the screen recordings and developed time-stamped logs of participants' canvas interactions using spreadsheets. 
The recordings of the post-task interviews were transcribed, and the same coders independently analyzed the transcripts using a separate spreadsheet application.
As the unit of analysis~\cite{suthers2010framework}, we used each individual action on the canvas for the video-based analysis and participants' utterances for the interviews.
The codes were consolidated and refined across eight sessions over two months. 
The researchers used six of these sessions to deliberate, resolve conflicts around the interpretation of data, and revise the codes, and two sessions with the full research team informed the development of the three themes.
The result of these deliberative sessions was an iteratively developed codebook, which the first author used to conduct a final round of coding across the dataset.

The researchers' prior experience with collaborative ideation tools might have drawn attention to spatial organization.
We considered this perspective when interpreting the canvas actions alongside participants' discussions and interview accounts.
We therefore present the themes as situated interpretations of these seven groups' interactions (Table ~\ref{tab:themes}), recognizing that other researchers may interpret the same actions differently~\cite{braun2019reflecting}.
We did not calculate statistical inter-coder reliability because the codes were developed iteratively throughout the analysis. Instead, we used negotiated agreement~\cite{mcdonald2019reliability}, in which the researchers compared their independently documented interpretations and discussed ambiguities to refine the evolving codebook and clarify the themes.

\section{Findings}
Our analysis of canvas interactions and post-task interview responses revealed three themes (Table~\ref{tab:themes}) around how participants used artifact placement and arrangement as a coordination resource to signal intent and participate without verbal exchange, how these contribution patterns gave rise to a central and peripheral spatial logic on the canvas without explicit negotiation, and how these repeated patterns of contribution and artifact work made divisions of labor visible over time, from which collaborative roles emerged without explicit assignment.
These themes reveal how spatial organization in an infinite canvas can shape participation, coordination, and role differentiation during remote collaborative ideation.

\begin{figure}[th]
    \centering
    \includegraphics[width=\linewidth]{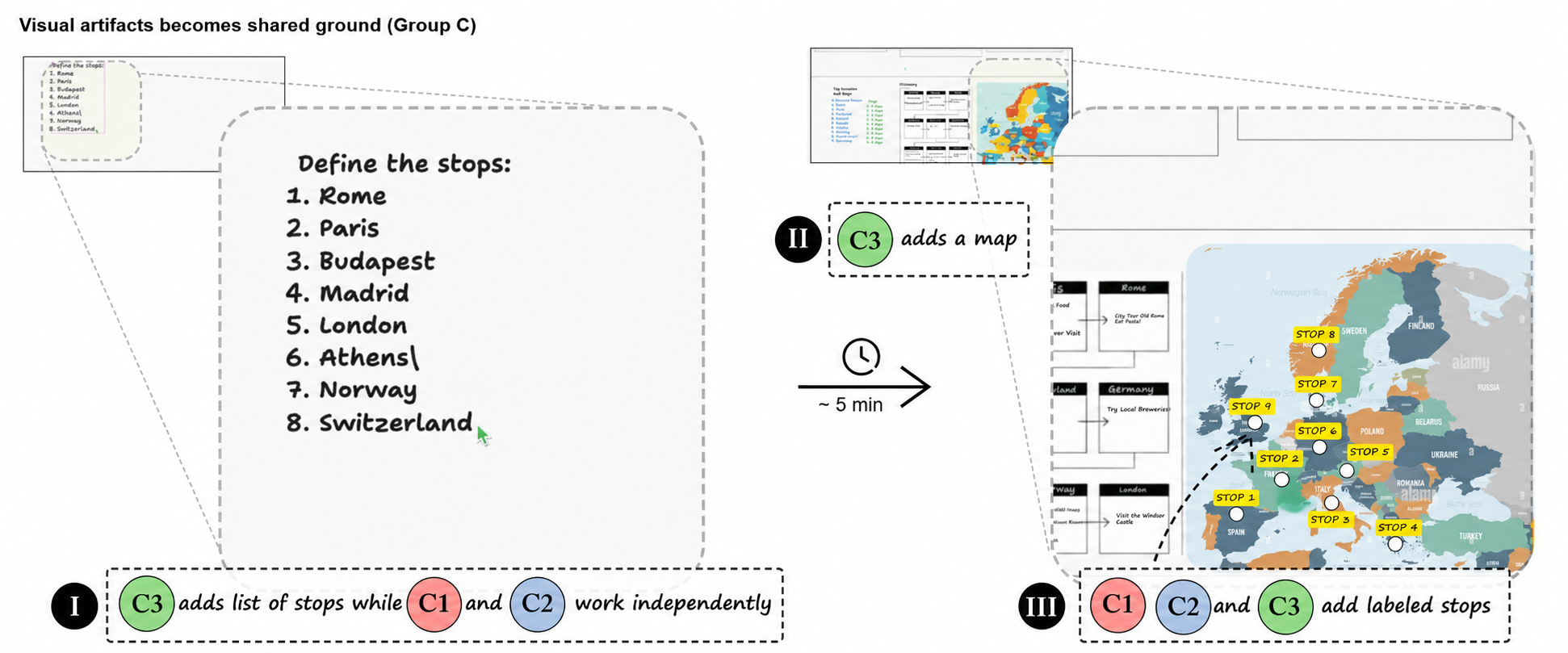}
    \caption{
    % (A) In Group F, participants used spatial organization to communicate relationships among ideas on the canvas. F1, F2, and F3 initially combined notes and images, with arrows linking artifacts to show connections between places and activities. As the session continued, later additions followed the emerging structure rather than being placed independently. Related artifacts became clustered together, and arrows helped represent chronological flow. 
    \circled{I} In Group C, visual artifacts evolved into a shared planning reference for the group. \circled{C3} first added a written list of stops while \circled{C1} and \circled{C2} worked on other things. \circled{II} Shortly afterward, \circled{C3} added a map, which became the anchor for collaborative planning. \circled{III} The group then began adding labeled stops to the map, using it to coordinate discussion and organize the itinerary spatially.
    }
    \Description{Three-panel storyboard from Group C showing how a shared planning reference emerged. In the first panel, C3 adds a written list of stops while the other participants work elsewhere. In the second panel, C3 adds a map that becomes the visual anchor of the group’s planning. In the third panel, the group adds labeled stops to the map and uses it to coordinate and organize the itinerary spatially.}
    \label{fig:storyboard1}
\end{figure}

\subsection{Artifact Organization Supports Implicit Coordination and Quiet Participation} \label{sec:4.1}

Participants spatially arranged multimedia artifacts to communicate relationships among ideas, signal priorities, and guide the next steps in the planning process without explicitly stating their intent. 
Across groups, participants added notes beside or overlapped with images to annotate them.
They clustered and connected artifacts using arrows and links to indicate shared themes or order.
Moreover, when they contributed new ideas, those artifacts were often added in line with the existing arrangements, continuing the groups' organizational patterns, without much verbal discussion.
For example, in Group C, participant C3 first added a written list of stops while C1 and C2 worked on different ideas.
Then, C3 added a map of the continent and annotated it with labeled stops and day markers, which became a shared planning region on the canvas. Later, C1 and C2 began contributing to the map in the same manner, without deliberate discussion (Fig. \ref{fig:storyboard1}).
Reflecting on this during the interview, C2 explained, \emph{``Seeing how others had arranged content made it easier to know where to add mine,''} suggesting that participants read the existing layout as a guide for where and how to contribute.

\begin{figure}[th]
    \centering
    \includegraphics[width=\linewidth]{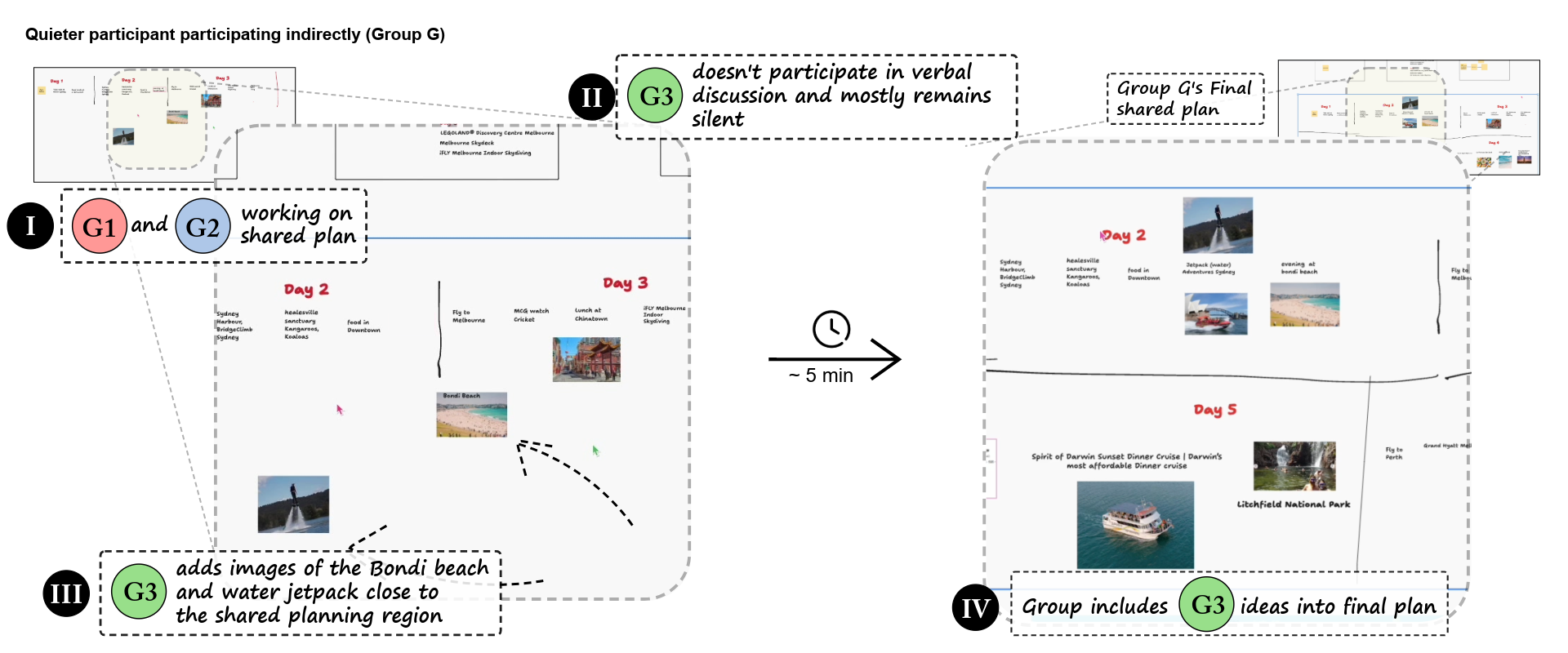}
    \caption{ \circled{I} In Group G, while \circled{G1} and \circled{G2} work on a shared day-to-day plan together, \circled{II} \circled{G3} remains mostly silent and does not communicate verbally during the period. \circled{III} But, \circled{G3}, the quieter participant, contributes by adding pictures of the Bondi beach (a destination spot) and water jetpack (an adventure sport) close to group's planning region on the canvas. \circled{IV} Other participants acknowledged the suggestions and those contributions were taken up into the developing plan, and \circled{G3}’s ideas became part of the group’s final shared plan.
    }
    \Description{Four-panel storyboard from Group G showing how a quieter participant contributed through proximity-based artifact placement. In the first panel, two participants work on a shared day-by-day itinerary. In the second panel, G3 remains mostly silent. In the third panel, G3 adds images of Bondi Beach and a water jetpack close to the active planning region. In the fourth panel, the group takes up these contributions and incorporates them into the final shared plan.}
    \label{fig:storyboard-fig4}
\end{figure}

Spatial organization also supported participation for collaborators who engaged less verbally during the planning session.
Instead of participating through verbal discussion, we observed participants with lower verbal engagement, including A3, B2, D2, and G3, contributing by placing artifacts close to the group’s active planning region, making their ideas visible within the developing layout without direct verbal assertion.
In Group G, for example, while G1 and G2 worked on the shared day-by-day itinerary, G3 remained mostly silent and added images of Bondi Beach and a water jetpack near the group’s planning area (Fig.~\ref{fig:storyboard-fig4}).
G1 and G2 saw those contributions, acknowledged them and even incorporated them into the developing itinerary which became part of the group’s final shared plan. Reflecting on this, G1 said, \emph{``We didn't think of any water adventure sports, but when G3 brought in those images, we liked it, and included them in our itinerary.''}

\begin{figure}[th]
    \centering
    \includegraphics[width=\linewidth]{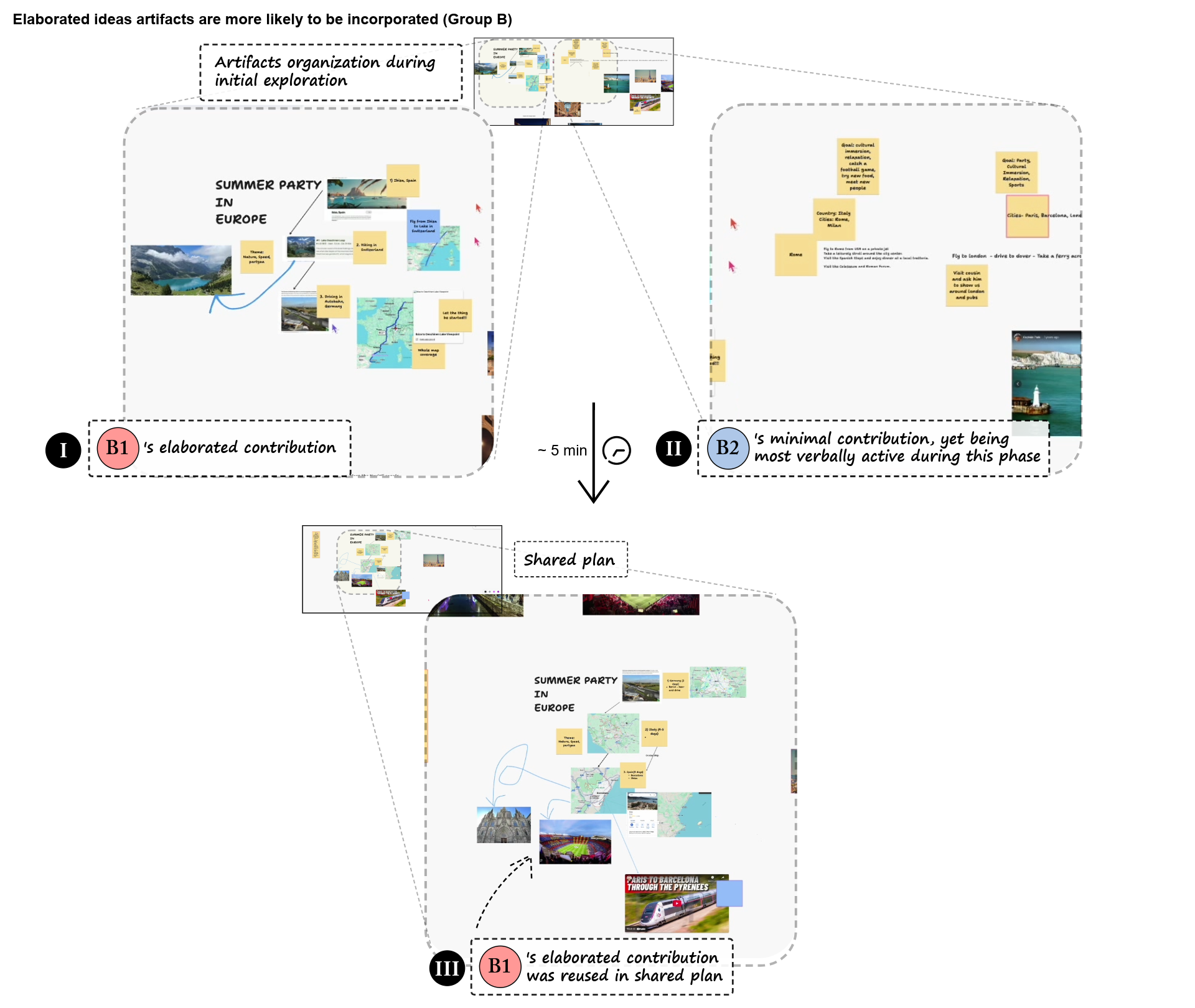}
    \caption{ \circled{I} During the initial exploration phase in Group B, \circled{B1} created a more elaborated arrangement using images, notes, and visible structure, \circled{II} while \circled{B2}’s on-canvas contribution remained minimal despite being more verbally active during the collaborative phase. \circled{III} As the session progressed, the shared plan continued to follow \circled{B1}’s organizational pattern, and \circled{B1}’s elaborated contribution was carried forward into the final shared canvas.
    }
    \Description{Three-panel storyboard from Group B showing selective uptake of a more elaborated contribution. In the first panel, B1 creates a detailed arrangement with images, notes, and visible structure. In the second panel, B2’s on-canvas contribution remains minimal despite being verbally active. In the third panel, the group’s shared plan follows B1’s earlier organizational pattern, and B1’s elaborated contribution is carried into the final shared canvas.}
    \label{fig:storyboard-fig3}
\end{figure}

However, not all contributions were equally taken up into the final shared plan.
Contributions that included images or annotations, or that were organized in a visually distinct way, were more likely to be read and taken up by others, while sparse text-only contributions were often abandoned early or required further development before being incorporated. 
This pattern appeared to stem from two related mechanisms. 
First, visual contributions prompted engagement and reconsideration in participants. A2 reflected on this, \emph{``There were list of Banff National park hike and trip to Mexico as suggestion. But later, someone added visuals of what it looked like in Miami, and I had to reconsider my choice... seeing the picture, I realized it's actually okay to go to Miami.''}
Second, contributions organized using arrows, labels, or grouping functioned as reusable templates that others could extend instead of replacing. 
This was visible in Group B, where participant B1 developed a detailed individual layout using images, captioned notes, and directional arrows, while B2 contributed minimally to the canvas. Despite B2 being more verbally active during the collaborative phase, collaborators used B1’s arrangement as the foundation for the shared plan and also maintained its organizational pattern throughout the session (Fig. \ref{fig:storyboard-fig3}).

These observations show that artifact organization on the canvas carried communicative value beyond presentation. 
The spatial arrangement of content functioned as a quiet but consequential coordination resource through which participants signaled intent, interpreted each other's actions, and found ways to join the collaboration and shape the shared plan.

\begin{figure}[h]
    \centering
    \includegraphics[width=\linewidth]{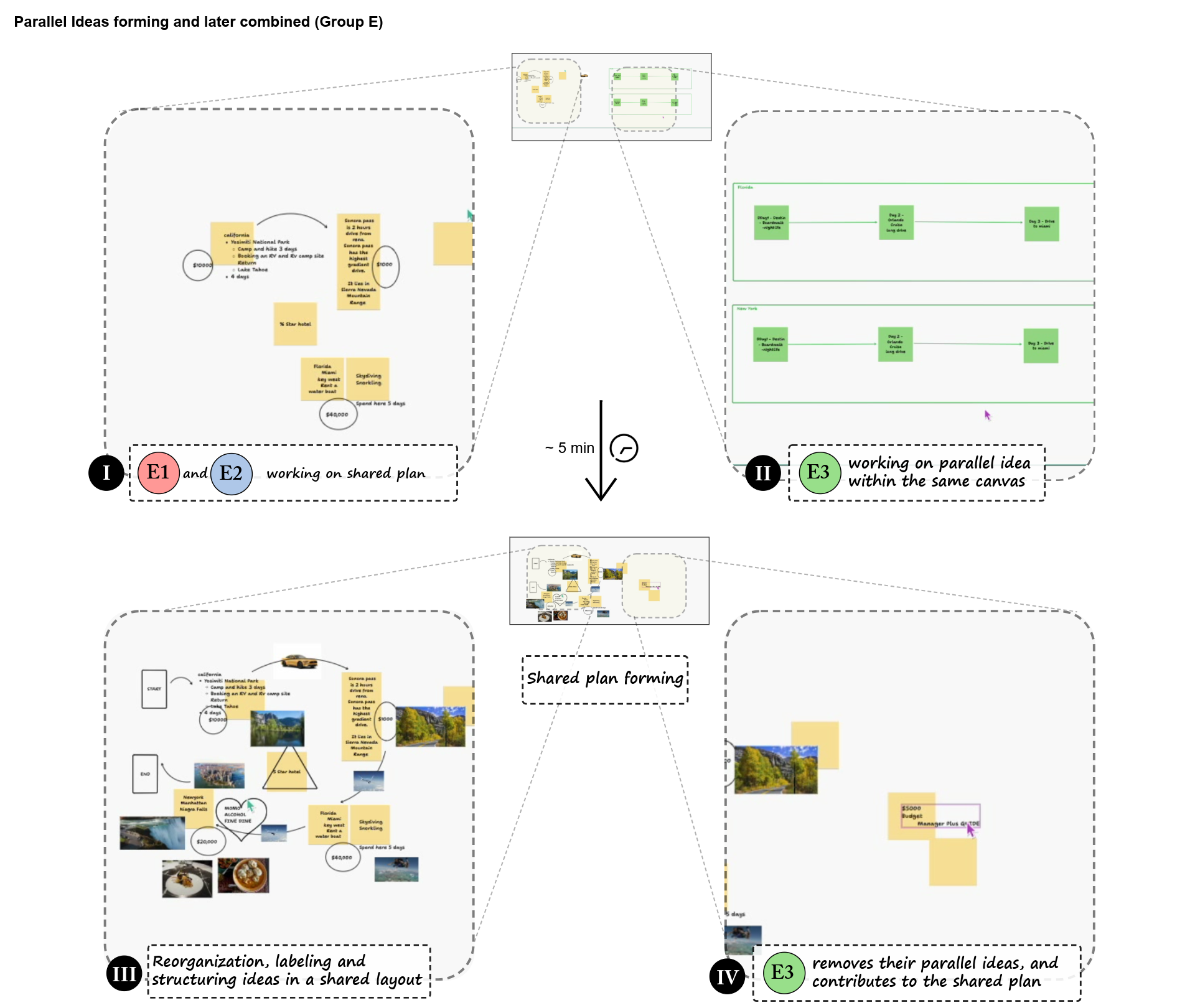}
    \caption{ \circled{I} In Group E, \circled{E1} and \circled{E2} work on the shared plan, \circled{II} while \circled{E3} develops a parallel idea in a separate area of the canvas. \circled{III} As the session progresses, the group reorganizes, labels, and arranges artifacts into a shared layout, and \circled{IV} \circled{E3} removes their separate parallel idea and contributes to the emerging shared plan.}
    \Description{Four-panel storyboard from Group E showing how shared spatial structure emerged over time. In the first panel, E1 and E2 work on a shared planning region. In the second panel, E3 develops a separate parallel idea in another part of the canvas. In the third panel, the group reorganizes, labels, and arranges artifacts into a more unified layout. In the fourth panel, E3 removes the separate idea region and contributes to the emerging shared plan.}
    \label{fig:storyboard-Fig5}
\end{figure}

\subsection{Contribution Patterns Led to Spatial Differentiation of Canvas Without Explicit Negotiation} \label{sec:4.2}

Even without predefined structure, participants developed a shared spatial logic on the canvas over time through artifact placement and organization.
The canvas imposed no boundaries on where ideas could be placed, so participants added contributions wherever space allowed. Early in the collaborative session, in 5 out of 7 groups, participants began by placing artifacts across the canvas with weak grouping, developing multiple local idea regions in parallel.
As the session progressed, participants reorganized initially scattered contributions through various organization schemes, such as clustering---grouping artifacts in proximity to each other, alignment---placing relevant artifacts in rows, columns, or overlapping, and labeling---adding text such as titles or captions to groups of artifacts.
In Group E, for instance, E1 and E2 initially worked on organizing artifacts together while E3 explored a separate idea thread elsewhere on the canvas (Fig.~\ref{fig:storyboard-Fig5}). 
Over time, participants clustered these parallel idea threads together into a common space on the canvas through reorganizing and labeling, eventually forming a more unified shared layout.

\begin{figure}[h]
    \centering
    \includegraphics[width=\linewidth]{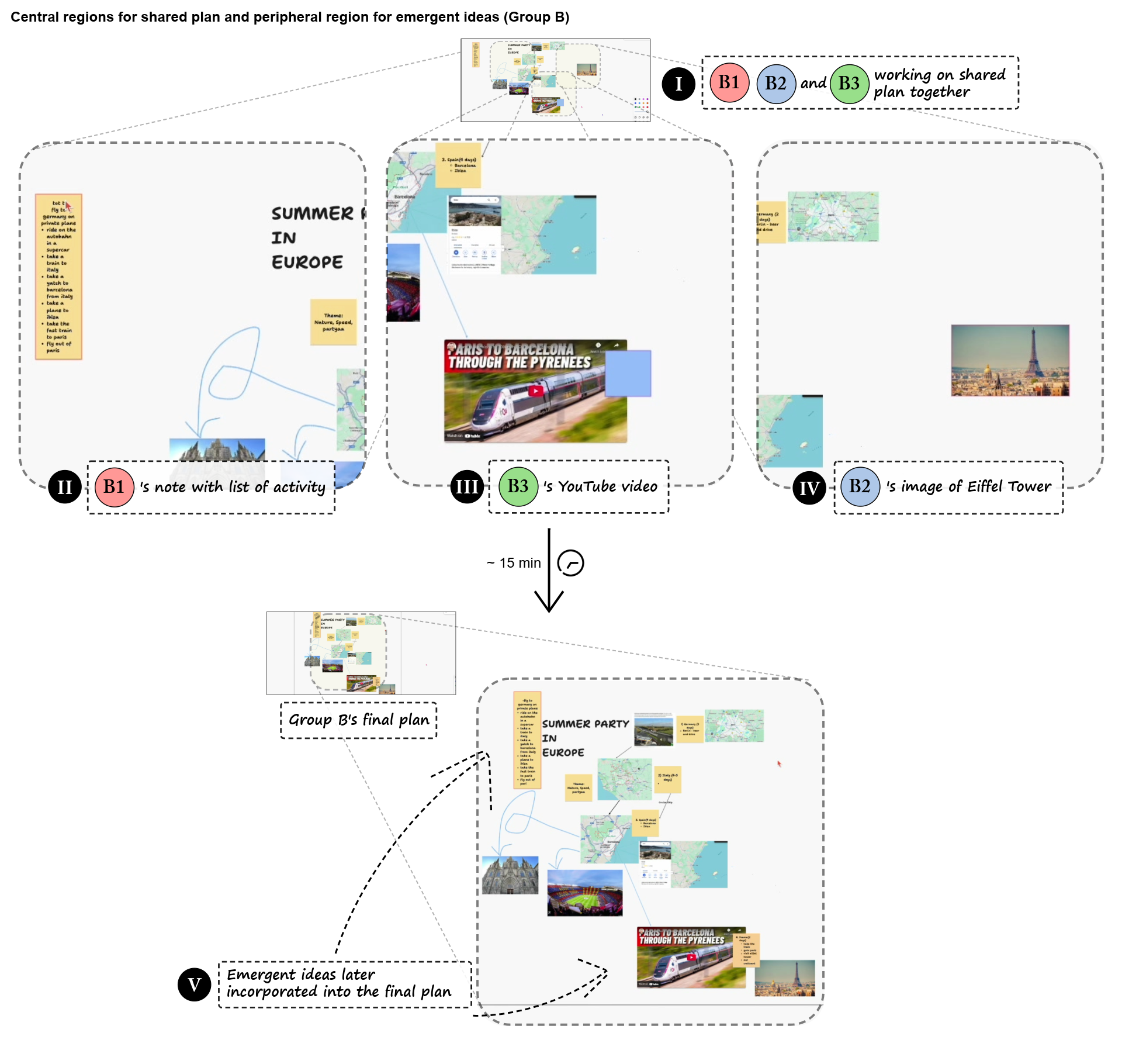}
    \caption{ 
     \circled{I} In Group B, while \circled{B1}, \circled{B2} and \circled{B3} are working on a shared plan together, some ideas were first introduced at separate regions. \circled{II} \circled{B1} adds a note with list of activities on the left hand side spatially separate from shared region, \circled{III} \circled{B3} brings in a YouTube video of a highly recommended train trip in Europe, \circled{IV} \circled{B2} places an image of Eiffel tower on a separate space on the canvas. \circled{V} The group later incorporates these ideas into the plan after they were further developed,  or became relevant to the group’s ongoing work.
    }
    \Description{Four-panel storyboard from Group B showing differentiation between shared and peripheral canvas regions. In the first three panel, the group works together on a shared plan while some ideas remain separate. Individual ideas are added outside the central working region, including a note listing activities, a YouTube video about a train trip in Europe, and an image of the Eiffel Tower. In the final panel, some of these peripheral ideas are later incorporated into the shared plan when they become more developed or relevant.}
    \label{fig:storyboard-Fig6}
\end{figure}

Participants also described rearranging the artifacts to compare and evaluate ideas. D1 explained that, \emph{``placing ideas visually allowed the group to put each of our alternative ideas side by side and examine what is similar and what is overlapping.''}
Through these patterns of contribution and comparison, the canvas became differentiated into central regions where shared and negotiated work accumulated, and peripheral spaces where tentative or not-yet-incorporated ideas remained.
We observed that this pattern of placement gave rise to a functional spatial logic on the canvas.
Group B's collaboration illustrated this progression (Fig.~\ref{fig:storyboard-Fig6}).
The group's shared canvas developed in one central working area, where the group actively consolidated their plan together. 
In parallel, individual artifacts remained in spatially separate regions, including B1’s note listing activities, B3’s video, and B2's images, placed away from the shared planning canvas region. 
None of these were introduced into the central plan immediately.
Over time, some of these artifacts were brought into the shared region when they became more developed or relevant to the group's ongoing work.

These observations suggest that the spatial differentiation observed across groups was a consequence of how participants contributed and participated, even without deliberate planning or agreement.
Over time, a spatial logic emerged on the canvas that distinguished shared ideas from more tentative and exploratory ones.

\begin{figure}[th]
    \centering
    \includegraphics[width=\linewidth]{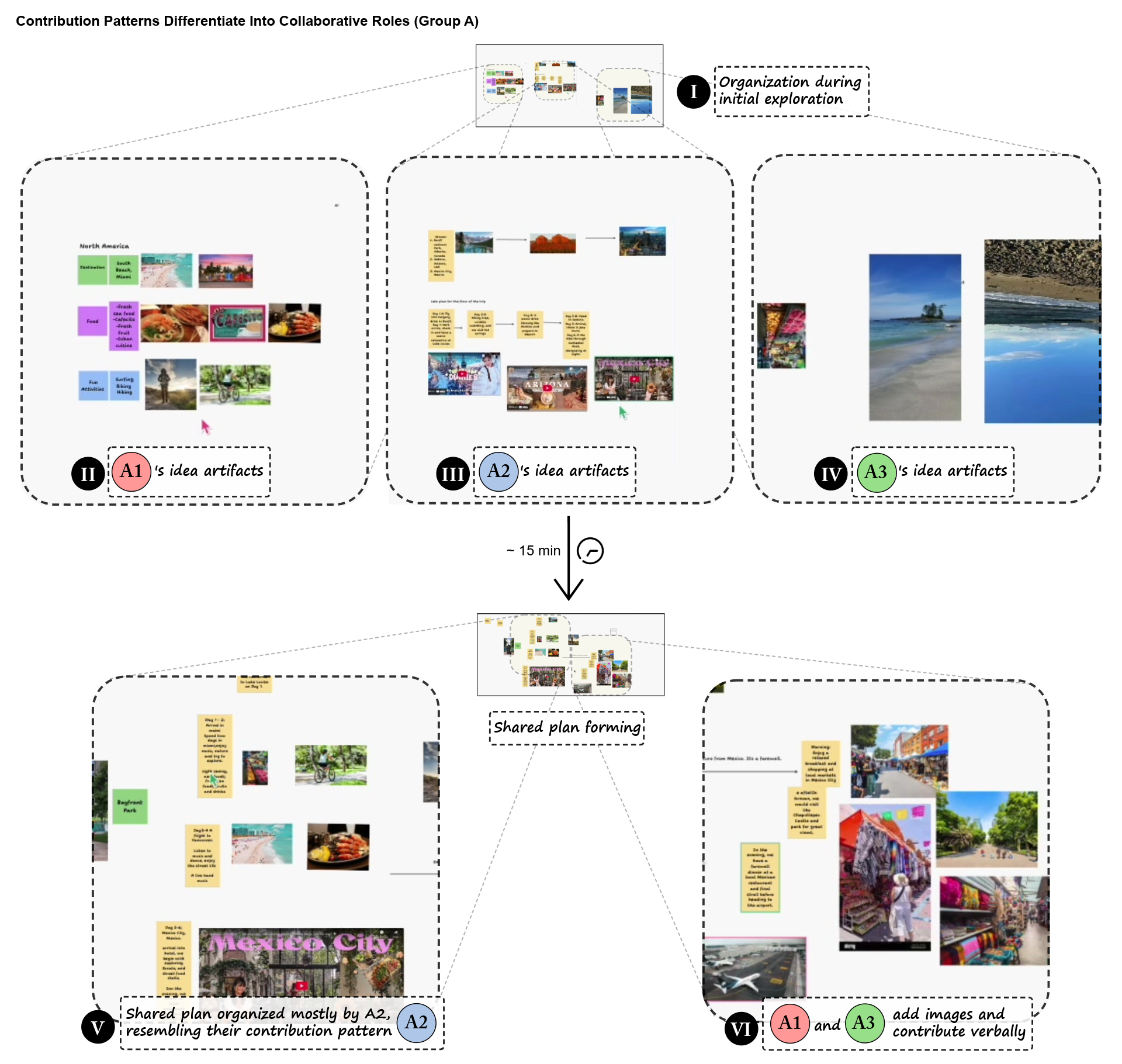}
    \caption{Role emergence in Group A. \circled{I} During the initial exploration phase, \circled{A1}'s, \circled{A2}'s, and \circled{A3}'s organization of contributions were visibly different on the canvas. \circled{II} \circled{A1} arranged notes and images in stacked rows, \circled{III} \circled{A2} arranged notes and images connected with arrows and included some YouTube videos, \circled{IV} \circled{A3} contributed images of destination spots. \circled{V} As the shared plan begins to form, \circled{A2} took charge of arranging the artifacts on the canvas. The group’s organization follows closely from \circled{A2}’s earlier contribution pattern, \circled{VI} while \circled{A1} and \circled{A3} brought in images and contributed verbally.}
    \Description{Six-panel storyboard from Group A showing emergence of collaborative roles through repeated patterns of contribution. Early panels show different contribution styles: A1 arranges notes and images in stacked rows, A2 creates a more structured composition with arrows and videos, and A3 contributes a smaller set of destination images. In later panels, as the shared plan develops, A2 takes charge of organizing the canvas while the other participants contribute images and verbal input.}
    \label{fig:storyboard3}
\end{figure}

\subsection{Roles Emerge Through Persistent Patterns of Contribution on the Canvas}
\begin{figure}[th]
    \centering
    \includegraphics[width=\linewidth]{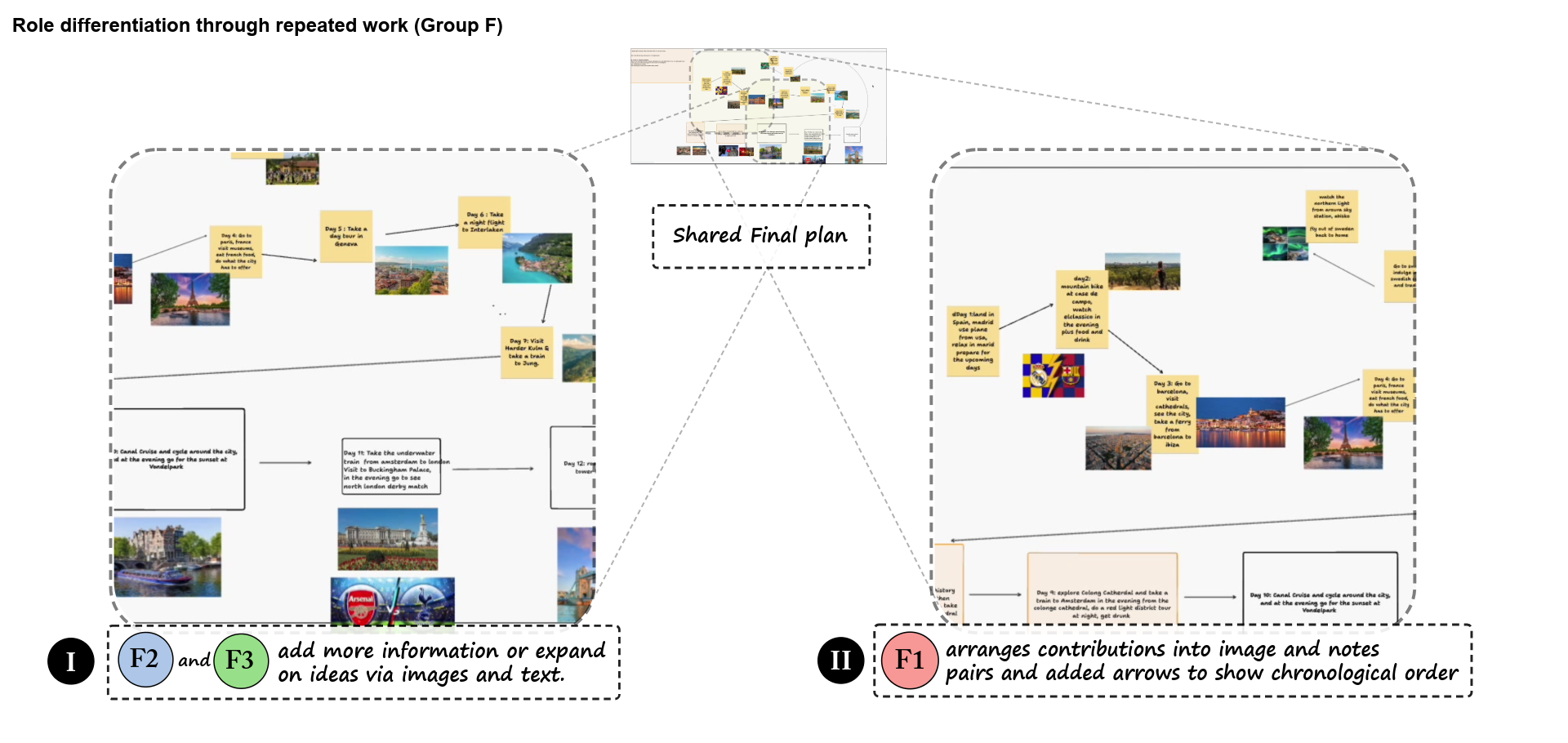}
    \caption{Role differentiation in Group F. \circled{I} \circled{F2} and \circled{F3} expand on ideas by adding images and text. \circled{II} \circled{F1} arranges contributions into image-and-note pairs and adds arrows to show chronological order, consolidating the group's individual contributions into a shared final plan.}
    \Description{Two-panel figure from Group F showing role differentiation. In the first panel, F2 and F3 add images and text that expand on existing ideas. In the second panel, F1 organizes the group's contributions into paired image-and-note groupings connected by arrows showing chronological order, which feed into the group's shared final plan.}
    \label{fig:storyboard9}
\end{figure}

We observed that participants gradually took on different collaborative roles through repeated patterns of contribution rather than through explicit assignment. Some participants mostly organized artifacts, while some added content.
In 5 of the 7 groups, one participant emerged as the primary canvas organizer, responsible for arranging, grouping, and maintaining the shared layout, while other members expanded on ideas by adding specific artifacts, such as images, web content, or list.
Also, some participants contributed primarily through verbal discussions while making fewer direct edits on the canvas. 
These contribution behaviors remained visible over time on the canvas, and division of labor gradually emerged during the session.
For instance, in Group A, A1 arranged notes and images in aligned rows, A3 contributed only a few images, and A2 produced a composition with notes and images, connected artifacts with arrows, and incorporated videos.
As the session shifted into collaborative planning, A2 took on the role of primary canvas organizer, while A1 and A3 contributed more through discussion and occasional edits, as illustrated in Fig.~\ref{fig:storyboard3}.
A similar pattern of role emergence was discussed in Group E's interview. 
E3's sustained organizing work led E3 to take charge of the plan, while E1 helped coordinate discussion and E2 remained mostly verbal. 
E1 reflected on this shift: \emph{``[The shared planning process] didn't go smoothly for the first few minutes, until we realized what each person can do, who can be that person who can organize. After a while, we found each other's strength and assumed responsibility, then I think it went smoothly.''}
The same pattern appeared in Group F, where F1 arranged the group's contributions into image-and-note pairs and added arrows to show chronological order, while F2 and F3 expanded on ideas by adding images and text (Fig.~\ref{fig:storyboard9}).

Moreover, we observed that organizing work was also linked to broader awareness of the shared plan. 
Across all seven groups, the participant who did the most organizing work also volunteered to present the final plan, including A2, B3, C3, D1, E3, F1, and G1.
Interview responses also suggested that these participants were more likely to recall and explain how the plan had developed, compared to teammates who were less involved in structuring the shared workspace. 
In Group G, for example, G1, who actively organized the canvas, noted during the final presentation that \emph{``I could explain [our plan] without referring to the canvas and [that'd be pretty close to] what already outlined in the canvas.''} suggesting that those who built the idea and layout on the canvas could communicate the plan in its totality with ease.
We also observed that organizers sometimes took on leadership-like responsibilities by helping direct the flow of work and clarify how things are organized. 
F3, who contributed by bringing in images of places, reflected, \emph{``F1 suggested to me what we should add next, and how to add them and I could just help F1.''}

These observations suggest that role differentiation emerged through sustained interaction with artifacts rather than prior agreement. 
The canvas made the division of collaborative labor readable as work unfolded, while repeated structuring activity appeared to foster both coordination and a more integrated awareness of the overall plan.

\section{Discussion}

In this study, we examined how groups in remote settings used an infinite canvas to collaboratively ideate. 
Drawing on our findings, we discuss what these interaction patterns and experiences suggest about coordination, participation, and role emergence in canvas-based collaboration, and implications for the design of future tools.

\subsection{The Canvas as a Communication Surface: Spatial Action as an Alternative Participation Medium}

Our findings suggest that the infinite canvas provided less vocal participants with a pathway into the collaboration that did not depend on securing the conversational floor. 
Rather than waiting for a pause or explicitly asking to be heard, these participants placed artifacts near the group’s active planning region. 
As the session progressed, collaborators noticed, discussed, and incorporated many of these contributions into the shared plan, even when their authors remained relatively quiet. 
Although persistent shared materials can support contribution in other settings, the canvas's open layout allowed participants to communicate tentative ideas by placing artifacts near ongoing work without fitting them into a fixed sequence.
This allowed ideas to remain peripheral or provisional until others were ready to engage with them, creating a form of delayed participation that did not require interrupting the ongoing conversation.

These observations extend prior research on workspace awareness and collaborative work showing that participation in shared environments does not depend on explicit verbal negotiation, but can instead emerge through actions made visible to other collaborators~\cite{dourish1992awareness,gutwin1996workspace,kwok2024unobtrusive}. 
Kirshenbaum et al.~\cite{Kirshenbaum2021Traces} report a related pattern on shared large displays: placed content persists over time rather than disappearing as speech does, and proximity between related ideas signals their connection to other collaborators. 
The canvas in our study did not impose a fixed structure, so participants could express intent and priority simply through where and how they placed artifacts, without stating it aloud. 
This appears to be the mechanism behind quiet participation in our data: an artifact, once placed, remained visible to the rest of the group, and its position relative to the group's active work conveyed relevance on its own, even absent the gaze, gesture, and proximity cues a co-located group would ordinarily rely on. 
Shared text documents leave far less room for this kind of participation, since contributions there are bound to a linear order and fixed formatting~\cite{han2020textlets}.

However, relying on visibility and proximity alone can also work against quiet participation. If more contributions accumulate on the canvas, or if a group includes several quiet participants at once, noticing and interpreting each placed contribution becomes harder for everyone else, and some contributions may go unseen. 
Future work should explore ways to support quiet participation without placing this burden entirely on other collaborators. For example, canvas tools might surface unnoticed contributions for group attention, or use subtle awareness cues to flag new placements near an active region~\cite{lee2021collabally, coppens2024awareness}.

\subsection{Designing for Implicit Coordination and Emergent Role Awareness}

Our findings show that the infinite canvas supported diverse contributions from group members, which gradually coalesced into roles, even though no roles were assigned at the outset. 
Participants made different contributions, where some arranged content or kept the layout coherent, while others extended existing structures, collected new material, or evaluated and annotated ideas.
As these actions repeated over time, they made unspoken roles visible. 
Notably, the participants who spoke the most did not always lead the collaboration. 
Instead, influence came from sustained organizational work on the canvas. 
Those who structured and maintained the shared layout often shaped how the plan developed and how others contributed to it.

These observations resonate with prior work in Computer Supported Cooperative Work (CSCW) showing that coordination in shared environments can occur through visible actions, which go beyond explicit verbal negotiations~\cite{dourish2001action,gutwin1996workspace,kwok2024unobtrusive}.
Others suggest that UX professionals increasingly use infinite-canvas tools such as Miro~\footnote{\href{https://miro.com/}{https://miro.com/}} and Figjam~\cite{figjam} to develop and maintain shared awareness of each other's contributions through visible artifacts and annotations~\cite{feng2023UX}. 
Because the canvas itself does not impose a fixed structure, participants could express their artifact organization through actions. 
We argue that, during collaboration, as more visible actions accumulate over time, the organizational contributions and competence become visible, enabling the emergence of roles through repeated patterns of contribution. 
Similar role dynamics can occur in shared text documents, where organizational work follows a more linear structure~\cite{lee2021collabally, han2020textlets}. 
In our study, the open canvas made emerging responsibilities visible through participants' repeated arrangement and maintenance of shared artifacts.

However, the lack of organizational structure in infinite canvases invites contradiction to the emergence of such roles. 
For example, if participants' contributions are spread across large areas of the canvas, this may lead to disorientation~\cite{das2024career}, as tracking contributions becomes more challenging and their visibility is affected. 
Further investigation is needed to develop alternative approaches for supporting emergent role patterns without turning them into rigid assignments. 
For instance, infinite-canvas tools might rotate structural stewardship, surface neglected or orphaned content for collective attention, or make organizational edits more legible through subtle awareness cues 
~\cite{lee2021collabally, coppens2024awareness}.

\subsection{Spatial Organization as a Design Resource for Future Support Tools}

Across our findings,  participants used spatial organization as an active resource to coordinate, participate, and signal intent (Sec.~\ref{sec:4.1}).
This observation points to a design direction where future tools could treat spatial layout as a signal.
Prior systems, for instance, IdeaHound, demonstrate this feasibility by inferring relationships between ideas from how participants arrange them~\cite{pao2016ideahound}.
Extending this idea, such design implications become more significant as AI facilitating agents enter the collaborative ideation tools.
This raises questions around ways in which AI agents could operate on canvases where meaning is carried by placement as well as the content.
The recent surge of AI has already permeated the collaborative ideation tool design space, with recent works demonstrating how AI agents can support human effort in spatial and relational structuring to high-level idea curation~\cite{yun2026modeling, he2024ai, rayan2024genai}, and researchers have begun exploring how AI might be integrated into collaborative ideation~\cite{Shin2023Integrating, rosenberg2024drawtalking, lin2025inkspire}. 
However, like many other areas that value human ingenuity, scholarship on the collaborative ideation space also promotes caution in integrating AI agents into collaborative workflows without careful consideration of human agency and control~\cite{wang2025aideation, Wadinambiarachchi2024genai, liu2024smart, sun2024wild}. For instance, Endsley~\cite{Endsley2023Supporting} shows that adding AI teammates to human teams often reduces coordination and trust, particularly when AI actions are opaque to other collaborators.

In our study, coordination also depended on participants being able to read each other's spatial actions.
AI agents that add artifacts or reorganize layout without making their reasoning visible risk disrupting this coordination mechanism, replacing legible human action with opaque automated change.
This concern aligns with established guidelines for human-AI interaction~\cite{Amerish2019Guidelines}, which emphasize that AI actions should be visible, understandable, and dismissible.
Leveraging the spatial affordances of infinite canvases, one potential approach to designing AI-assisted collaborative ideation tools could be to explore the separation of spaces for humans and AI agents. 
This aligns with prior work by Webb et al.~\cite{webb2011implicit}, who propose central workspaces, \textit{``hot spaces''}, where human users actively construct and refine ideas, while peripheral regions, \textit{``cool space''}, are used by automated agents to gather, suggest, and organize information in the background.

Our findings suggest that future AI-supported canvases should preserve the spatial cues through which collaborators interpreted one another's contributions. 
For example, automated additions or layout changes could be clearly attributed to the system and left for participants to accept or revise, consistent with prior work on transparency and human control in human-AI interaction~\cite{Endsley2023Supporting, Amerish2019Guidelines}.

\section{Limitations and Future Work}

Our findings should be interpreted in light of following limitations. 
First, our study involved seven triads (N = 7$x$3 = 21). Our findings are grounded at the group level, and having only seven groups might constrict the scope of interpretation and generalization. 
Second, all groups consisted of three participants, and it remains an open question whether the coordination mechanisms, spatial structures, and role patterns we observed would hold in larger groups.
Third, our study focused on collaborative planning task in a remote setting and should not be interpreted as representative of workplace collaboration more broadly. Behaviors may differ in professional settings, domain-specific tasks, or longer-term collaborations.
Fourth, participants had different levels of experience with collaborative tools, and the study did not require prior experience with canvas-based tools. 
Participants' familiarity with canvases may have influenced how they organized the canvas and used its features.
Our study does not establish how prior canvas experience shaped these behaviors.
Finally, our study used a custom technology probe rather than a commercial tool such as Miro, which allowed for detailed interaction logging but means some behaviors may reflect the specific affordances of the probe.

Future work could study open-canvas collaboration in larger groups, longer-term collaborations, and a wider range of task types such as brainstorming, sensemaking, or design work, where spatial organization may function differently.
% {It would also be valuable to study open-canvas collaboration in a wider range of tasks beyond planning, such as brainstorming, sensemaking, or design work, where spatial organization may function differently. }
Another direction is to investigate how systems might better support the transition of ideas across central and peripheral regions, make quiet forms of participation more visible, or help collaborators recognize emergent roles as they develop. 
More broadly, future research could explore how the spatial and multimodal properties of open canvases can be designed to better support awareness, coordination, and inclusive participation in both remote and hybrid collaboration.

\section{Conclusion}
In this work, we examined how remote groups collaborate in a multimodal infinite-canvas, and how ideas were contributed, organized, and incorporated into shared work over time. 
In a study with seven triads, we observed that collaboration on the canvas unfolded through discussion and ongoing interaction with shared artifacts. Over time, spatial organization emerged without predefined constraints, with central shared regions and peripheral spaces for tentative ideas, taking on different functions as ideas were revisited, clustered, and negotiated. 
We also found that the placement and organization of artifacts supported implicit coordination, helping collaborators interpret each other’s actions and determine where and how to contribute. 
Repeated forms of artifact work further made division of labor visible, as participants gradually took on different roles through arranging, extending, and maintaining shared content. 
These findings show how collaboration on an infinite-canvas develops through spatial, material, and social interaction, and they point to opportunities for future systems that better support shared awareness, coordination, participation, and emergent roles in collaborative ideation.

\bibliographystyle{ACM-Reference-Format}
\bibliography{references}

% \appendix









\end{document}